\documentclass{aastex63}

\received{June 1, 2019}
\revised{January 10, 2019}
\accepted{\today}
\submitjournal{ApJ}

\shorttitle{V3890 Sgr with Chandra}
\shortauthors{Orio et al.}

\begin{document}

\title{Chandra High Energy Transmission Gratings Spectra of
 V3890 Sgr}

\correspondingauthor{Marina Orio}
\email{orio@astro.wisc.edu}

\author[0000-0003-1563-9803]{M. Orio}
\affiliation{INAF-Osservatorio di Padova, Vicolo Osservatorio 5,
35122 Padova, Italy}
\affiliation{Department of Astronomy, University of Wisconsin,
475 N. Charter Str., Madison WI 53704, USA}
\author[0000-0002-0210-2276]{J.J. Drake}
\affiliation{Harvard-Smithsonian Center for Astrophysics,
 60 Garden Street, Cambridge, MA 02138}
\author[0000-0003-0440-7193]{J.-U. Ness}
\affiliation{XMM-Newton Science Operations Center, European Space Astronomy Center,\\
Camino Bajo del Castillo s/n, Urb. Villafranca del
Castillo,
 28692 Villanueva de la Ca\~nada, Madrid, Spain}
\author[0000-0002-9356-1645]{E. Behar}
\affiliation{Department of Physics, Technion, Haifa, Israel}
\author[0000-0002-2647-4373]{G.J.M. Luna}
\affiliation{CONICET-Universidad de Buenos Aires, Instituto de Astronomia y Fisica del Espacio (IAFE),\\
 Av. Inte. G\"uiraldes 2620, C1428ZAA, Buenos Aires, Argentina}
\affiliation{Universidad de Buenos Aires, Facultad de Ciencias Exactas y Naturales, Buenos Aires, Argentina}
\affiliation{Universidad Nacional de Hurlingham, Tte. Origone 151, Hurlingham,
Buenos Aires, Argentina}
\author[0000-0003-0156-3377 ]{M.J. Darnley}
\affiliation{Astrophysics Research Institute, Liverpool John Moores University, Liverpool, L3 5RF, UK}
\author[0000-0001-8608-0408]{J. Gallagher}
\affiliation{Department of Astronomy, University of Wisconsin,
475 N. Charter Str., Madison WI 53704, USA}
\author[0000-0003-1319-4089]{R. D. Gehrz}
\affiliation{Minnesota Institute for Astrophysics, University of Minnesota, 116 Church Street S.E, Minneapolis, MN 55455,
 USA}
\author[0000-0003-4650-4186]{N.P.M. Kuin}
\affiliation{University College London, Mullard Space Science Laboratory, Holmbury St. Mary, Dorking RH5 6NT, UK}
\author[0000-0003-3457-0020]{J. Mikolajewska}
\affiliation{Nicolaus Copernicus Astronomical Center, Polish Academy of Sciences, Bartycka 18, PL-00-716 Warszawa, Poland}
\author[0000-0002-8404-1808]{N. Ospina}
\affiliation{Department of Physics and Astronomy,
 University of Padua, vicolo Osservatorio, 3, 35122 Padova, Italy}
\author[0000-0001-5624-2613]{K.L. Page}
\affiliation{School of Physics and Astronomy, University of Leicester,
University Road, Leicester LE1 7RH, UK}
\author[0000-0002-9968-2464]{R. Poggiani}
\affiliation{Dipartimento di Fisica, Universit\`a di Pisa, I-56127 Pisa, Italy}
\affiliation{INFN, Sezione di Pisa, I-56127 Pisa, Italy}
\author[0000-0002-1359-6312]{S. Starrfield}
\affiliation{Department of Earth and Space Exploration,
Arizona State University, PO Box 871404, Tempe, AZ 85287-1404}
\author[0000-0002-3742-8460]{R. Williams}
\affiliation{Department of Astronomy \& Astrophysics, University of California, Santa Cruz, 1156 High Street, Santa Cruz, CA 95064, USA; Space Telescope Science Institute, 3700 San Martin Drive, Baltimore, MD 21218, USA}
\author[0000-0001-6567-627X]{C. E. Woodward}
\affiliation{Minnesota Institute for Astrophysics, University of Minnesota, 116 Church Street S.E, Minneapolis, MN 55455,
 USA}
%
\begin{abstract}
The recurrent nova (RN) V3890 Sgr was observed during the 
 7th day after the onset of its most recent outburst, with the {\sl Chandra}
 ACIS-S
 camera and High Energy Transmission Gratings (HETG).
 A rich emission line spectrum was detected,
 due to transitions of Fe-L and K-shell ions ranging from neon to iron. 
 The measured absorbed flux is $\approx 10^{-10}$ erg cm$^{-2}$ s$^{-1}$
 in the 1.4-15 \AA \ range (0.77-8.86 keV). 
 The line profiles are asymmetric, blue-shifted and skewed
 towards the blue side, as if the 
ejecta moving towards us are less absorbed than the receding ones. 
The full width at half maximum of most emission lines is
1000-1200 km s$^{-1}$, with some extended blue wings.
The spectrum is thermal and consistent with a plasma in collisional
 ionization equilibrium with column density 1.3 $\times 10^{22}$ cm$^{-2}$
and at least two components at temperatures
 of about 1 keV and 4 keV, possibly
a forward and a reverse shock, or regions with differently
 mixed ejecta and red giant wind.
The spectrum is remarkably similar to the symbiotic RNe
 V745 Sco and RS Oph, but 
 we cannot distinguish whether the shocks occurred 
 at a distance of few  AU from the red giant,
or near the giant's photosphere, in a high density medium
 containing only a small mass.
The ratios of the flux in
 lines of aluminum, magnesium and neon relative to the flux in lines 
 of silicon and iron probably indicate a carbon-oxygen white dwarf (CO WD).
\end{abstract}

a

\section{Introduction} \label{sec:intro} 
 Recurrent Novae (RNe), like all novae, are interacting binaries with a 
 WD that accretes matter from its companion and
 undergoes a thermonuclear runaway. They are called ``recurrent'' because
 they have been observed
 in outburst more than once, although all novae are
 thought to have repeated outbursts on different, secular time scales 
\citep[e.g.][]{Prialnik1986}. Only a small number of RNe are known, 
10 in the Galaxy \citep[see review by][]{Orio2015}, 3 in 
 the Large Magellanic Cloud \citep{Mroz2016, Bode2016}
 with a recent additional candidate \citep{Ilkiewicz2019}, 12 in M31 with
 4 additional candidates \citep{Shafter2015}. Of the 10 known Galactic RNe,
 4 are ``symbiotic novae'', relatively rare novae in long orbital period
 systems with a red giant companion.

 V3890 Sgr is one of the symbiotic RNe, 
  and it was observed in outburst for the
 third time on 2019-08-27.87 \citep[it was reported
 by A. Pereira, see][]{Strader2019}. Previous outbursts
 occurred in 1962 and 1990, spaced 28 years apart.
 The last quiescent optical photometric
observation was made less than a day before the outburst
 detection \citep{Strader2019}, so the outburst time is fairly well constrained.
Table 1 summarizes the known physical parameters for V3890
 Sgr.  The current outburst
 of V3890 Sgr may have been faster, with shorter t$_2$ and t$_3$
 (times to drop 2 and
 and 3 magnitudes from
 maximum optical luminosity, respectively), than the previous
 one, according
 to the AAVSO publicly available light
 curve. However, it is possible that 
 the initial phases in the previous eruptions of 1962 and 1990 were missed,
 so that t$_2$ and t$_3$ may not
 have been well constrained.
 The time to decay by 6 magnitudes was about the same
 in each outburst \citep[][and references therein]{Strope2010}.

 For a useful comparison, Table 1 also includes the known parameters
 for  the other three Galactic symbiotic RNe, 
 which have been observed and scrutinized
 in the past much more extensively than V3890 Sgr, especially RS Oph.
We note that all four objects have an M spectral type companion and are
 fast novae with short t$_2$ and t$_3$. We also know that
 the orbital periods exceed one year in RS Oph and  V745 Sco.  
 Symbiotic RNe are somewhat outliers among novae in several ways. 
 The WD mass, m(WD), has been estimated to be very
 high for RS Oph, V745 Sco and T CrB.
 As for most RNe, the maximum absolute optical magnitude cannot
 be empirically related to the time for a decay of 2 or 3 magnitudes via
 the maximum-magnitude-rate-of-decay relationship \citep[MMRD; see][]{DellaValle1995}. 
 The ejected material in V3890 Sgr, as well in the
 other symbiotic RNe in Table 1,
 had high velocity, reaching at least 
 4000 km s $^{-1}$. this  fact is not explained yet  by the models of the 
 thermonuclear runaway (TNR), which indicate slower ejecta than
 observed for several RNe, both symbiotic and otherwise  \citep{Yaron2005}.
  The short interoutburst times imply 
  high accretion rates, $\dot m \geq 10^{-8}$ M$_\odot$ yr$^{-1}$,
 necessary  to build enough pressure at the base
 of the envelope and  start the TNR again every few
 years or tens of years \citep[e.g.][]{Yaron2005}.  
 RS Oph and T CrB are thought to be disk accretors, like 
 generally many other symbiotics, although this does not imply 
 that all symbiotics with a disk fill their Roche Lobe.
 In fact, most symbiotics do not
 fill it,  although several of them, never observed
 in nova outburst, show signatures of semi-steady nuclear burning
\citep[see][]{Orio2007, Orio2013, Mikolajewska2012}
  requiring even higher $\dot m$ than in RNe.
 Such high $\dot m$ is not reached with Bondi-Hoyle accretion \citep{Bondi1944},
 yet in a few cases the systems' parameters are known with sufficient precision
 to rule out Roche Lobe filling.
 Often ellipsoidal variations are observed,
implying that 
some symbiotics are tidally distorted and for
 this reason some material indeed flows 
 through the inner Lagrangian point L1 \citep[see][and references therein]{Mikolajewska2012}. 
In other symbiotics, including the well studied 
 RS Oph, no tidal variations are observed, yet there
 are signatures of a disk. \citet{Mohamed2013}
suggested that the stellar wind of
 the RS Oph giant flows through the L1 point because it is gravitationally focused
 and flows in a spiral shock on the equatorial plane. 

 Because of the circumstellar red giant wind, the nova ejecta in symbiotic novae impact
 colder and slowly moving material (with velocity
 of the order of about only 10 km s$^{-1}$),
 causing strong shocks, that are observed in X-rays. 
   Like other symbiotic novae, symbiotic RNe should also emit significant
 gamma-ray flux, as observed in the first nova detected
 with Fermi-LAT, V407 Cyg, which is also a symbiotic \citep{Abdo2010}. 
 However, V745 Sco, the first symbiotic RN in outburst in the Fermi era,
 was only marginally detected with Fermi \citep[see][and
 references and discussion therein]{Cheung2014, Orio2015b}.

 The new outburst of V3890 Sgr has been monitored at all wavelengths
 with many different
 instruments, also because RNe with high mass WDs (see Table 1) are interesting as possible progenitors 
 of SNe Ia while they are still ``single degenerates'' \citep[e.g.][]{Starrfield2019}.
 Early optical and near infrared spectra have been described, among others, by
\citet{Strader2019, Evans2019, Mehara2019, Munari2019a,
Munari2019b, Pavana2019a, Rudy2019, Woodward2019}.  
 An ultraviolet spectrum was obtained by \citet{Kuin2019} and
 {\it Swift} also provided the UV light curve.
 In Fig. 1 we present the ``early'' {\sl Swift} X-Ray Telescope (XRT) light
 curve.
 Hard X-rays were immediately detected with the {\sl Swift} XRT
 \citep{Sokolovsky2019} and the nova was a gamma ray source \citep{Buson2019},
 as expected. 
 Already on the second day after the eruption, the Swift X-ray spectrum 
 showed a hard X-ray source that could be fitted as a thermal plasma with kT=7.9$^{+2.3}_{-1.6}$
 keV \citep{Sokolovsky2019}. 
No connection can be made between the measured X-ray emission and the gamma ray emission,
  since also some non-thermal X-ray flux is expected when gamma-rays
 are observed \citep{Vurm2018} and  only a thermal spectrum
 was observed with {\sl Swift} by \citet{Sokolovsky2019}.  
A radio detection was announced by \citet{Pavana2019b} with likely thermal
 emission, but also non-thermal,
synchrotron emission was later detected \citep{Pavana2019b, Polisensky2019},
 consistent with the gamma-ray detection.

 \citet{Munari2019c} obtained the reddening to the nova from the interstellar
 features in the optical spectra, deriving
 E(B-V)=0.56, and note that the comparison with
 interstellar reddening maps indicates a distance to the nova in
 excess of 4.5 kpc.
 
In this article, we present the {\sl Chandra} High Energy Transmission Gratings'
 (HETG) X-ray spectrum observed on the 7th day after the onset of the 
 optical outburst, as illustrated
 in Fig. 1, which allowed us to observe the hard X-ray emission from 
 the ejecta. 
In Fig. 1 we indicate the date of the {\sl Chandra} observation,
 to place it in the context of the nova evolution.
 We note that the hydrogen burning WD was observed as a luminous
 supersoft X-ray source (SSS) only shortly later, 8.4 days after the 
 outburst on 2019 September 5.25, when the ejecta had 
 become transparent to soft X-rays \citep{Page2019}. We note that 
 an additional high resolution X-ray spectrum was obtained with {\sl XMM-Newton}
 for the supersoft X-ray phase \citep{Ness2019} and is the subject
 of another article, currently being prepared.

 In Section 2 we describe the data we obtained; Section 3 contains an analysis
 of the spectrum
with proposed models to fit the data; in Section 4 we compare
 the {\sl Chandra} HETG spectra observed for two of the other
 symbiotic RNe and for a classical nova, V959 Mon. They were
 all observed at a similar evolutionary post-outburst
 phase. We also briefly discuss what 
the upper limits on quiescent X-ray emission imply for the cooling time
 of the ejecta and the secular evolution of the binary.
 Section 5 contains our conclusions.
\begin{figure}
\plotone{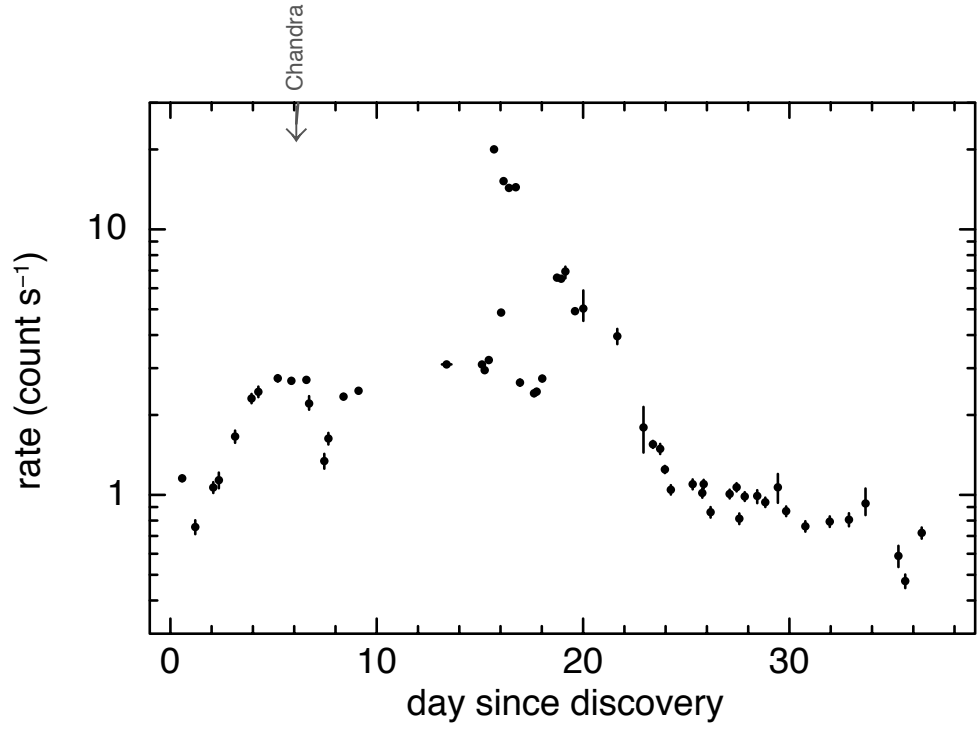} 
\caption{The light curve of V3890 Sgr measured with the {\sl Swift} XRT 
 in the 0.5-10 keV range. The
 date of the {\sl Chandra} observation is indicated by the arrow.
 An XRT light curve measured over a long period is included in a paper
 by Page et al. 2020, in preparation.}
\label{fig:swift}
\end{figure}
\begin{longrotatetable}
 \begin{deluxetable*}{lrrrrrrrrrrrrl}
\tablecaption{Physical parameters of the known symbiotic recurrent novae.
 The distances
 without reference are obtained from the GAIA DR2 parallax with calculation of statistical
 error from the {\it gaia.ari.uni-heidelberg.de/TAP} web site,
 but we caution that the long orbital period of symbiotics 
 may have introduced errors
larger than estimated in the parallax obtained up to the data release
 DR2, thus, 
 additional distance estimate are reported.  
 The times for optical
 decay by 2, 3 and 6 magnitudes are from \citet{Strope2010}, and the second
 row for V3890 Sgr reports our measurement
 in the AAVSO light curve in the 2019 outburst (www.aavso.org).
 The ``times(SSS)'' are the times,
 in days., for the SSS to emerge, to remain at maximum
 luminosity, and to fade beyond detection, respectively. \label{conf}}
\tablewidth{1100pt}
\tabletypesize{\scriptsize}
\tablehead{
\colhead{Name} & \colhead{m(WD)} & \colhead{m(giant)} &
\colhead{spectral} & \colhead{d} & \colhead{P$_{\rm orb}$} & \colhead{t(rec)} &  \colhead{V$_{max}$} & \colhead{v(ej) } & \colhead{t$_2$}  & \colhead{t$_3$}  & \colhead{t$_6$}  & \colhead{times(SSS)} \\
\colhead{}     & \colhead{M$_\odot$} & \colhead{M$_\odot$} &  \colhead{type} &
\colhead{kpc} & \colhead{days}   & \colhead{years}          & \colhead{}  & \colhead{km s$^{-1}$} & \colhead{days} & \colhead{days} & \colhead{days} & \colhead{days}   
}
\startdata
 V3890 Sgr  &          &           & M5 III (1,2)  & 4.36$^{+2.64}_{-1.31}$ & (see 3) & 28       & 7 & $\geq$ 4200 (4)  & 6 & 14 & 28 & 8-20-26 (**)  \\   
            &          &           &               & $\geq$4.5 (5)   &        &           &   &      & 2 & 5  &  27  & \\ 
 RS Oph     & 1.2-1.4 (6)  & 0.68-0.80 (6) & M0-2 III (6) & 2.29$^{+0.28}_{-0.27}$ & 453.6 (6) & $\approx$10(*) & 4.8 & 4200 (7) & 7 & 14 & 88 & 26-58-86 (8) \\  
            &              &               &              & 1.6 (9)                &           &  & & & \\
 V745 Sco   & $>$1.3 (10) &          & M6$\pm$2 III (1) & 7.28$^{+4.52}_{-2.92}$ &  & 26? & 9.6 (1)  & $\geq$4000 (11) & 2 & 4 &  & 4-6-7 (12) \\ 
            &           &                  & & 7.8$\pm$1.8 (10) &       &             &  &            & & & & \\  
 T CrB      & 1.2$\pm$0.2 (13) & 0.7 (13) & M4 III (14) & 0.822$^{+0.034}_{-0.076}$ & 227.67 (15) & 80  & 2 &  & 4 & 6 & 15 &  \\   
            & 1.37$\pm$0.13 (14) & 1.12$\pm$0.23 (16) & & & & & & & & & & \\
\enddata
\tablecomments{ (1) \citet{Harrison1993}; (2) \citet{Anupama1999}; 
 (3) \citet{Mroz2014} - These authors find that the proposed
 orbital periods of V3890 Sgr and V745 Sco in \citet{Schaefer2010} cannot
 be confirmed; (4) \citet{Strader2019}; 
 (6) \citet{Brandi2009};
 (7) \citet{Buil2006} ; (8) \citet{Bode1987} and references therein 
 (9) \citet{Osborne2011}; (10) \citet{Shara2018};
(11) \citet{Banerjee2014}; (12) \citet{Page2015}; 
(13) \citet{Belczynski1998}; (14) \citep{Murset1999}; (15) \citet{Lines1988}; (16) \citet{Stanishev2004}.\\ (*) However, the last outburst of RS Oph occurred only after 21 years and no new outburst has
 been recorded in the last 13.5 years.\\
(**) Page et al. 2020, in preparation }
\end{deluxetable*}
\end{longrotatetable}
\section{The observed spectrum}
The {\sl Chandra} exposure of V3890 Sgr was performed on 2019 September 03
 (starting 6.4 
 days after the initial detection at optical wavelength) 
with the ACIS-S camera and the High Energy Transmission
Gratings (HETG). Both the medium energy grating MEG and the high
energy grating HEG were used, with a respective absolute wavelength accuracy of 0.0006 \AA \
and 0.011 \AA. The exposure time was 30 ks, 
and we measured a count rate  0.2139$\pm$0.0027 cts s$^{-1}$ in the zeroth order ACIS-S camera.
 Half of the
incident radiation is dispersed to the gratings in this observation mode, and the 
count rates were 0.5378$\pm$0.0061 cts s$^{-1}$
 in the HEG summed first orders (energy range 0.8-10 keV), and 1.0450$\pm$0.0063
 in the MEG summed first orders (energy range 0.4-10 keV).
We extracted the spectra with the CIAO software \citep{Fruscione2006}
 version 4.9.1 and the CALDB calibration package version 4.8.3. We
 integrated the flux in the ranges where most
 of the flux is and the value of S/N is high,
 obtaining  1.01 $\times 10^{-10}$ erg cm$^{-2}$ s$^{-1}$
 in the 1.5-14 \AA \ range (0.89--8.27 keV) for the HEG, 
 and 1.03 $\times 10^{-10}$ erg cm$^{-2}$ s$^{-1}$ in the 1.8-16 \AA \ (0.75--6.89 keV) range for the MEG.
 Spectral fitting was done with the HEASOFT XSPEC tool, version 12.6,
 after the data were binned by signal-to-noise with the GRPPHA tool
\citep[see][and references therein]{Dorman2001}.

 The HEG and MEG spectra are shown in Fig. 2; the prominent emission lines
 are marked on the HEG spectrum. 
 Here we adopted the count rate on y-axis for comparison with \citet{Drake2016}
 and other papers with HETG figures in the literature that use these units.
The two spectra are in perfect
 agreement when plotted with units of flux in the y-axis,
 which we use in the next figures. 
We notice prominent H-like and He-like lines of elements
 ranging from Ne to Fe, and some additional Fe-L shell lines.

\section{Spectral fit and interpretation}
 A thermal plasma model with a single temperature is not 
sufficient for a statistically acceptable fit, but
 Fig. 2 shows also
 the best, and statistically acceptable, spectral fit obtained
 with two BVAPEC components of thermal plasma in collisional ionization
 equilibrium (CIE) \citep[Astrophysical
 Plasma Emission Code, or APEC, with variable abundances
 and velocity broadening in XSPEC, see][and references therein]{Smith2001}.
 We used the TBABS prescription for the absorbing, equivalent column density
 of hydrogen atoms \citep{Wilms2000} and the solar abundances  
 listed in \citet{Asplund2009}. 
 We found that $\chi^2$ statistics were suitable to assess the goodness of the
 fit.
 The parameters of two models and the reduced $\chi^2$ value
 are given in Table 2. In both
 cases, the lines components are modeled as blue-shifted by several
 hundred km s$^{-1}$ and broadened.  We constrained the
 broadening velocity of the cooler component not to exceed
 that of the hotter one, because we assume that the broadening
 indicates turbulence.  
 We note that this constraint yields the same best fit $\chi^2$ value 
 as in the case the broadening velocity is allowed to vary
 independently for the two components (and the broadening velocity
 of the cooler component turns out to be larger). In
 the first model, we also constrained all abundances to solar values.
  The two components of the fit are 
 at 0.99 keV and 3.99 keV, respectively.
 The column density is N(H)= 1.32 $\times 10^{22}$ cm$^{-3}$.
The total flux is overestimated in this fit by about
 5\%, and the unabsorbed flux at the source is calculated
 to be 2.62 $\times 10^{-10}$ erg cm$^2$ s$^{-1}$.
This fit, shown in Fig. 2, underestimates the flux in several lines, especially
 He-like ones, albeit by a small amount.  However, we could not obtain a 
better fit by adding more components at different temperature.

 In the second model in Table 2, we kept two
 components, but allowed all the abundances of elements
 with significant emission lines in the spectrum, that is  Ne, Mg, Al, Si, S,
 Ar, Ca and Fe, to vary independently from each other,
and also independently for each component.  
 However, the cooler component, which may contribute much less than
 the hot one  to the continuum, but has to account for at least
 four emission lines for each element (H-like, and He-like triplets),
 is largely unconstrained  in the fit with all
 these parameters. The 90\% statistical
 uncertainties of the emission measure and abundances 
 are very large without any a-priori assumptions;
 our data are not sufficiently good for so many free parameters. 
 Nevertheless, in Table 2 we also give the best fit parameters we obtained with
 this second model, again constraining the broadening velocity
 of the cooler component not to exceed that of the hotter
 component. 
 The flux was overestimated by  almost 8\%, but we obtained  a lower 
 value of $\chi^2$ per degrees of freedom. 
 The two components are at temperatures
close to those of the previous case,   0.97 keV and 4.06 keV respectively.
  This best fit is
 obtained with very enhanced abundances for the cooler
 component, and very high absorption. We obtain 
 instead about nearly solar abundances for the hotter
 component, except for Al, which turns out to be very depleted. 
 The best fit value of N(H) is lower than in the previous
 model, 8.3 $\times 10^{21}$ cm$^{-3}$, 
 and the unabsorbed flux is 2.04  $\times 10^{-10}$ erg cm$^{-2}$ s$^{-1}$.
  We stress, however, that given the large uncertainties,
 the only clear conclusion is that  the fit
 improves if we assume that the cool component has enhanced
 metal abundances, and that the hot one is consistent with near-solar values.
\begin{figure}
\plotone{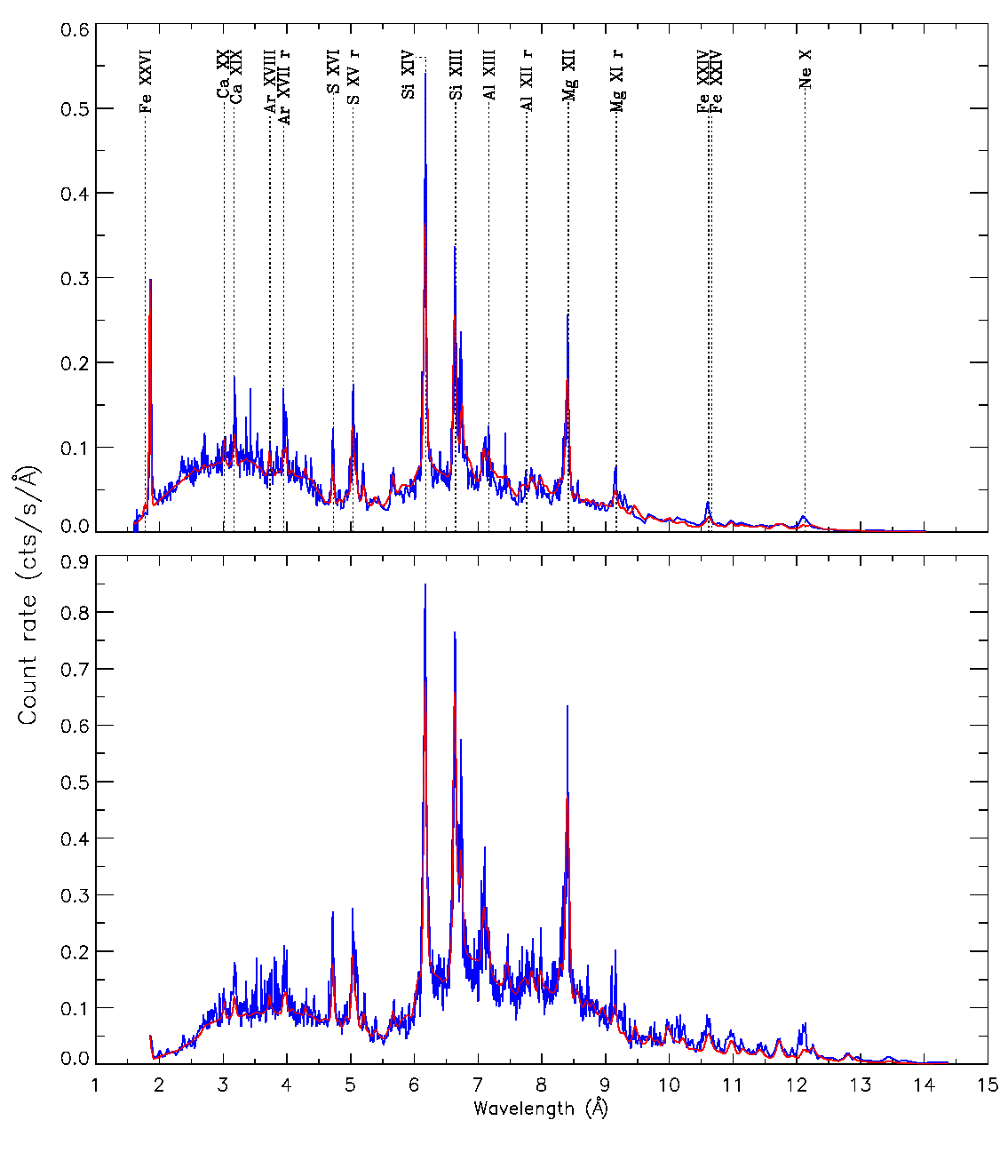}
\caption{The HEG spectrum is plotted in the upper panel and the MEG in the lower one. Both
 show wavelength as function of count rate and the strong lines are marked in the
 HEG panel. The best fit with two APEC thermal components (Model 1)
is traced in red in both
 panels.}
\label{fig:hegandmeg}
\end{figure}

We came to the conclusion that a CIE is a suitable model by examining 
helium-like triplets in the spectrum. 
For each ion, the flux in the ``r'' recombination line,  in
 the ``i'' intercombination line and ``f'' forbidden
 line gives the so called G ratio, $G=(f + i)/r$, which indicates
whether the plasma is in CIE. Generally,
 G$>$4 indicates a contribution of photoionization \citep{Gabriel1969,
 Bautista2000, Porquet2010}
as long as this diagnostic is used in a regime where the
 forbidden line is not sensitive to the density,
that is, at extremely high densities \citep[compatible with values 
 of electron density n$_{\rm e}\geq10^{10}$ cm$^{-3}$ in this early
 post-maximum phase, a value that was inferred
 at optical wavelengths by,][among others]{Neff1978}.  
As Fig. 3 shows for the Si triplet, the
 He-like lines were detected with relatively low signal-to-noise,
 so that the intercombination line is barely observable. It also
 overlaps with the r and f line, hindering a  precise measurement. 
 We thus estimated
 the $G$ ratio with a fit of three Gaussians to the three lines,
  tieing their broadening velocity to the same value.
  An example is shown in Fig. 3.
 The resulting flux estimates and their statistical uncertainties are reported
 in Table 3.  Using the flux values reported in the Table, 
 we calculated 
 $G$=0.94$^{+0.45}_{-0.38}$ for S XV, $G$=1.21$^{+0.16}_{-0.35}$
 for Si XIII,
  $G$=0.88$^{+0.34}_{-0.29}$ for Mg XI, and
 $G$=0.73$^{+0.89}_{-0.60}$ for Na X (the statistical uncertainty
 is given at the 90\% confidence  level).  
 Even if our uncertainties are relatively high,
 these $G$ values are all compatible with  a plasma in CIE.
\begin{figure}
\plotone{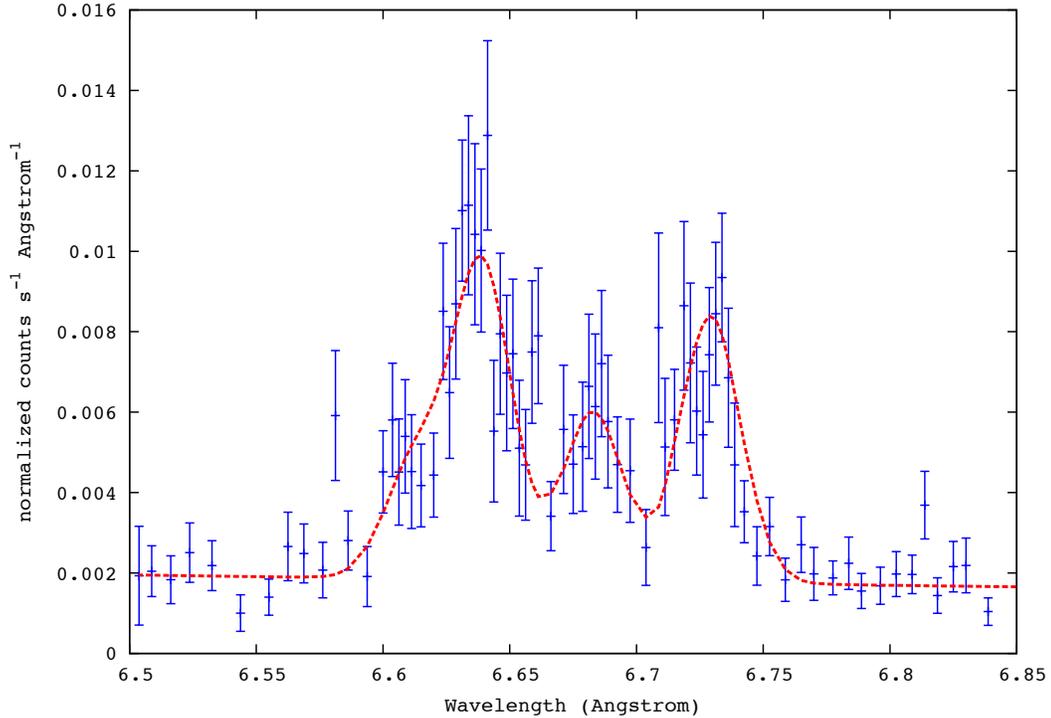}
\caption{The triplet of Si XIII plotted 
 binned by only 3, thus  bringing out the ``i'' line,
 and a possible fit with three Gaussians,  plus an additional line
 blending with the r line
 (this would be either a line of Al XIII at 6.63 \AA, or more
 likely, an additional
 high velocity component of the r line). The y axis is in units
 of photon flux. We did not fit here a
 likely additional emission line at $\approx$6.58
 \AA, sufficiently detached from the triplet. }
\label{fig:triplets}
\end{figure}

  The fluxes given in Table 3 are for lines 
 that could be measured with some precision, after subtracting the
 continuum flux from a nearby region with no emission lines.
 We can compare the fluxes with those obtained by \citet{Peretz2016}
 for V959 Mon, known as a nova on  an ONe WD. 
We find that the ratio of the flux of lines Al, Mg and Ne (elements
 expected to be enhanced in novae on ONe WDs), relative to 
 the flux of the Si lines, are much smaller than in V959 Mon, 
 despite plasma temperatures of about the same value.
 The ratio of the flux in the H-line of Si XIV to that of Al XIII
 is about 1.3 in V959 Mon, while is about 13 in V3890 Sgr. For Mg XII
 this ratio is about 1 in V959 Mon and 2.5 for V3890 Sgr
 (Table 1 of the above authors reports fluxes in units of
 incident photons $\times$ cm$^{-2}$ s$^{-1}$, and in order to
 compare them with our units of erg cm$^{-2}$ s$^{-1}$ they have to be 
 divided by the wavelength in \AA \ and multiplied by a
 factor of $\approx 2 \times 10^{-8}$). 
We suggest that these very different flux ratios are a strong
  indication in favor of classifying the WD in V3890 Sgr as a CO rather
 than an ONe one. This
 is corroborated by the outburst UV spectra of the symbiotic RN
 RS Oph (which is very similar to V3890 Sgr in many respects), that 
 also indicate a massive CO WD \citep{Mikolajewska2017}. 

 As Fig. 4 illustrates for four H-like lines,
 the lines we measure deviate from a simple
  Gaussian profile:
 they are asymmetric and skewed towards the blue.
 A possible interpretation is that the 
 red-shifted (receding material) is more absorbed than the
 blue-shifted material, which is moving towards the observer. 
The ejecta that are receding with respect to us 
 are observed through more foreground layers of ejected plasma. 
 Such an absorption effect is clearly observed in the emission lines
of supernova ejecta that form dust \citep{Gehrz1990}, 
 and was found also 
in the spectrum of RS Oph \citep{Nelson2008, Drake2009},
  V745 Sco \citep{Drake2016, Orlando2017} and V959 Mon,
 a classical nova with a large luminosity in ``hard'' X-rays,
 before the supersoft X-ray phase \citep{Peretz2016}. 
In the high resolution X-ray grating spectra of RS Oph the 
 apparent
 blue-shift was also found to increase with higher wavelength and lower ionization
 stage, as expected for absorption, which is stronger at  
 higher wavelength. Fig. 4 shows that this effect seems
 to be the same in V3890 Sgr, albeit less evident than in RS Oph. 

 Another major cause of the skewed profile,
 certainly overlapping with the differential
 absorption effect, is indicated by the global
 fit presented in Table 2.  In fact for the H-like lines, that have
 quite different profiles and strengths
 from the He-like lines, the  complex profile can be attributed 
 to at least two components moving at somewhat different velocity,
 and with different line broadening.  Consistently with
 the models in Table 2, we interpret this as due to one of
 the two  plasma components that
 is sufficiently hot to be almost completely ionized. 
 The hot component does not contribute
 to the flux of the He-like lines, as shown by the line profile and fit in
 the lower right panel of Fig. 5. Unlike the H-like lines, the 
 He-like triplets are fitted assuming
 only the contribution of the cooler component, although
 we remark that the global model
 more successfully reproduces the H-like lines than the
 the He-like ones. However, 
we had to bin the data by signal-to-noise of at least 20 in order to obtain
 a statistically significant fit in XSPEC, and with this
 binning the less strong ``i'' line
 appears smoothed out, so the global fit in XSPEC is done
 only using the ``r'' and ``f'' lines.  

 However, the ``erosion'' phenomenon due to differential absorption
 makes the models 
 with standard software like those in {\sl XSPEC} uncertain, because 
 the velocities of the two plasma components 
 cannot be precisely estimated without a detailed dynamical model 
 taking into account the effect of the intrinsic nova absorption
 in the ejecta along the line of sight
\citep[e.g. the hydrodynamical model of][]{Orlando2009, Orlando2017}.
Despite these uncertainties, it is quite clear that at least
 two components at different temperature are needed to explain the spectrum
 of V3890 Sgr, as in the X-ray spectra of
 the ejecta of other novae previously observed.
The first spectrum obtained for V959 Mon in 2012
 (albeit at a much later post-outburst epoch) was
 fitted with two plasma components at about the same temperatures 
as in Table 2, and the H-like lines
 indicated the contribution of a hot, almost
 fully ionized component at different velocity
 than a cooler plasma in which the He-like
 lines were formed. \citet{Mukai2019} have recently remarked
 that the two thermal plasma components that are suitable to fit the
 V959 Mon spectrum
 may be due to the reverse and forward shock; however 
 the data quality of \citet{Peretz2016}
 was not sufficient to estimate the velocity of
 the hotter component. It was consistent with a positive value, 
 versus the negative velocity of the cooler component. 
 
  Finally, there may be  an  undetected
 supersoft component in V3890 Sgr.  Two {\sl Swift} XRT exposures were
 performed during the Chandra observation; the first lasted only 450 s
 and the second 430 s, but the source was sufficiently bright for
 spectral fits.

%

\subsection{Possibility of a very clumpy medium} 
Another diagnostic that can be obtained from the He-like triplets is 
$R=f/i$, ratio of the fluxes in the $f$ and $i$ line of each triplet.
When this ratio
is low, the density is high enough that the collisional de-excitation 
rate from the upper level
 of the $f$ transition to the upper level of the $i$ transition
is competitive with the radiative decay rate in the $f$ line: the $f$
 line emission is thus reduced in favor of the $i$ transitions. 
 With our fit for the Si XIII triplet, we obtain a value $R=1.26^{+0.32}_{-0.33}$.
 We also find  $R$ values of 2.42$^{+1.57}_{-1.56}$ for S XV, and 0.69$^{+0.33}_{-0.28}$
 for Mg XI (the uncertainty represents the 90\% confidence level).
 For the CIE calculations pertinent to different databases,
 assuming a negligible radiation field, so that there is no strong UV flux
 causing radiative photoexcitation increasing the flux in the $i$ line,
 in Fig. 6 we show that
 the $R$ values we obtain for Si XIII and Mg XI, and  even their lower limits,
 indicate electron
 density higher than a few n$_e$  $\times 10^{13}$ cm$^{-3}$ for all triplets.
  While for the SXV triplet, the calculated $R$ ratio indicates 
 an electron density value above that examined  
 in the databases, the 90\%
 confidence level lower limit is consistent with  n$_e \geq 10^{14}$ cm$^{-3}$.
 We could not measure a significant value for Na X; the error bars are very large. 
 
Electron density above 
 10$^{13}$ cm$^{-3}$  appears very high compared
 with typical electron densities derived from optical spectra,
 which usually do not exceed 10$^9$ cm$^{-3}$ even at early phases.
 \citet{Neff1978} estimated a value n$_e = 1.7 \times 10^{10}$ cm$^{-3}$
 on the 9th day after the outburst for Nova Cyg 1975 (V1500 Cyg), but
 they found inhomogeneity in the ejecta
 and concluded they were measuring  upper limits for this and other
 parameters.  
\citet{Moore2012} calculated a post-shock electron density
 of the order of 10$^9$ cm$^{-3}$ for the post-shock material 
 in symbiotic recurrent novae 
{\it assuming a spherically symmetric expansion}. The models by
\citet{Orlando2009, Orlando2017} also imply post-shock electron density
of 10$^9-10^{10}$ cm$^{-3}$. 
However, we know that in two classical novae,  U Sco
 \citep{Orio2013} and V959 Mon \citep{Peretz2016},
 the high resolution X-ray spectra imply much higher density,
 at an epoch when the contribution of photoexcitation to the
 {\it i} line was already negligible. Whether this critical assumption
  on the photoexcitation contribution is also valid 
 for V3890 Sgr depends critically on the WD temperature on the day of
  observation, and on the distance at which the shocks occurred from
 the WD.  

According to model calculations 
 for the similar symbiotic nova RS Oph \citep[private communication by Prialnik,
 see also][]{Hillman2014} the  
 photospheric radius of the WD is about 3.5 $\times 10^{11}$ cm 
 during the whole period before the SSS emerges, then it
 shrinks to a value around 0.01 R$_\odot$ ($\approx 7 \times 10^{9}$ cm)
 within less than a day. The effective temperature before
 the shrinking leading to the SSS is about 30,000 K; it moves
 to values above 200,000 K within few hours.
 In order to assess the importance of photoexcitation
 we need to compare its rate with the decay rate of the 
 forbidden line. This is slow, and for Si XIII it is 3.5 $\times 10^5$ s$^{-1}$.
 The expression for the photoexcitation rate is 0.026 $\times f_{ij} F_{\nu(16eV)}$ where f$_{ij}$ = 0.054 is the oscillator strength for the
 transition between the upper levels of the forbidden and 
intercombination lines, and F$_{\nu(16eV)}$ is the UV flux
 at 16 eV at a given distance from the WD.  In our specific 
 case, the conclusion is uncertain because it is not straightforward
 to evaluate the photospheric effective temperature of the central source in 
 V3890 Sgr at the particular time of our {\sl Chandra} exposure.
 In fact, the WD was detected as an SSS only two days later, so 
 the WD radius may have already been shrinking, while the ejecta 
 may have still been largely absorbing the supersoft X-rays.
 There is a very large difference between 
 the luminosity at 16 eV of a blackbody-like source
 emitting 10$^{38}$ erg s$^{-1}$ with temperature of 30,000 K, about 
 about 10$^{37}$ erg s$^{-1}$,  and the luminosity at 16 eV of a source with  
 blackbody temperature above 200,000 K (as inferred with the {\sl Swift}
 XRT observation 2 days after our spectrum was taken). If our source
 was already at such a high temperature, while the ejecta were not yet quite
 transparent to the X-rays, the peak of the luminosity must have already 
 been in the X-rays and the 16 eV luminosity
 would have been negligible.

 If instead the effective temperature 
 during our observation was still 30,000 K and dropped only
 on the following day, the calculation indicates that the UV flux  was 
 still significant. In this case, the distance 
 at which the photoexcitation rate equals the decay rate of the Si XIII 
 forbidden line in astronomical units is 
 17.9 AU. In the model of \citet{Orlando2009}
 for RS Oph, the red giant is at 1.5 AU from the WD, and the peak of
 the flux is at 4 AU from the red giant, so the distance from
 the WD  would be inferior, only 5.5 AU. The photoexcitation rate 
 in such a case turns out to be about 10 times
 larger than the decay rate of the forbidden line. The contribution
 of the UV source at a distance of a few AU   
 would have  still been very significant and n$_e$ cannot be derived from
 the ratio of the triplets' lines.    

 It is interesting to note that for V745 Sco the $R$ ratios indicated 
 n$_e \leq 10^9$ cm$^{-3}$ \citep{Drake2009}, while
 in  RS Oph only the $R$ ratio derived from Mg XI may have been due 
 to electron density above 10$^{10}$ cm$^{-3}$, but the error
 bar was so large that the authors did not comment on it \citep{Nelson2008}. 
 In this respect, the X-ray spectrum of V3890 Sgr differs from the two other symbiotic RNe
 with similar orbital properties.
 If the WD photosphere had already moved towards
 a peak of  emission in the extreme UV and was not
 a UV source any more,  the high electron density we derive for V3890 Sgr is  
  comparable to the high values obtained for the classical novae
 U Sco and V959 Mon.

  What are the implications of
 an electron density of a few 10$^{13}$ cm$^{-3}$?
 One caveat is the very short radiative cooling time. 
 The ratio of the thermal energy of the plasma to the radiative loss
 rate (at the plasma temperatures we estimated with the spectral fit) imply a radiative
 cooling time of only 1 s, instead of several days as in
 the model of \citet{Moore2012}. This means that there must be a constant supply of
 ejected mass during all the time this thermal spectrum was measured,
 including the later and earlier {\sl Swift XRT} observations. 
Even with the maximum nova wind velocity estimated of about 4000 km s$^{-1}$,
 a spherically symmetric shocked region would be very thin, of order
 of only 4000 km. 

 The value of the total emission measure EM=$\int n_e n_{\mathrm i} \mathrm d V\approx n_e n_{\mathrm i} V$ that we obtain by adding the emission
 measure of the two model components in Table 2 is EM=1.81 $\times 10^{56} \times$ d(kpc)$^2$ cm$^{3}$
 in the solar abundance model and EM=9.76 $\times 10^{55}\times$ d(kpc)$^2$ cm$^{3}$
 in the model with variable abundances.  
 Assuming d=4.4 kpc, we find
 that EM=1.9--3.5 $\times 10^{57}$ cm$^{-3}$.
 If we assume that the X-ray emitting ejecta traveled at constant velocity for a week,
 and that they fill a spherical volume V, V=5.9 $\times 10^{43}$ cm$^3$, therefore
 our value of the emission measure yields n$_{\rm e} \approx 6-8 \times 
 10^6$ cm$^{-3}$, which is 6 orders of magnitude smaller than 
 estimated above using the Si XIII triplet,
 and cannot be reconciled with slower expansion
 velocity or a more refined calculation. We can only conclude that
 the volume filled by the shocked plasma is a very small fraction of
 the volume that homogeneous ejecta with a filling factor of 1 would fill. 

 Following
 the reasoning of \citet{Peretz2016} and, like these authors, assuming for
 simplicity a fully ionized gas with n$_e$=n$_p$ (where n$_p$ is
 the proton density) and approximately solar
 composition, the mass of the emitting plasma is 
\begin{equation}
 M=X Y~m_p~\frac {EM}{n_e}\approx11.25~m_p~\frac {EM}{n_e}
 \end{equation}   
 where $X$ is the mean proton number, $X=1.5$, and $Y$ is the mean atomic
 weight \citep[we assumed $Y=7.5$, see also][]{Peretz2016}.
 For $n_e \approx 10^{13}$ cm$^{-3}$,
 a simple calculation indicates that the emitting mass is $M=1.8 \times 10^{-12}$ M$_\odot$ assuming a 4.4 kpc distance,
 which is a very small fraction of M$_{\rm ej}=1.1 \times 10^{-6}$ M$_\odot$ 
estimated for instance for RS Oph by \citet{OBrien1992}  (we suggest that 
the total M$_{\rm ej}$ is 
likely to be of the same order for all symbiotic RNe, including
 V3890 Sgr \citep[see also model calculations by][]{Yaron2005}). 
This indicates that the X-ray emitting material is not uniformly
 distributed, but is rather concentrated in one small
 region, or in a number of clumps occupying a small volume.
 We note that,
 for densities of the order assumed by \citet{Orlando2009, Orlando2017}
 the shocked mass in this observation turns out to be instead of a few 10$^{-8}$ M$_\odot$. 
 Very clumpy emitting material
 was inferred for the non-symbiotic T Pyx \citep{Tofflemire2013},
 U Sco (Orio et al. 2013) and for V959 Mon \citep{Peretz2016}. In Section
 5, we discuss also the possibility that the shocks at the stage 
 at which we observed this symbiotic nova were occurring in
 a very small region, namely close to the the red giant.

\begin{deluxetable}{rr}
\tablecaption{Parameters of the best fits with two thermal components,
and reduced $\chi^2$ value.
 The abundance values are given in the text. The 90\% confidence level
 errors are given, but in the case of the variable abundances after
 they were calculated after having fixed the best fit abundances, and for the unabsorbed fluxes
 after having fixed also N(H).}
\tablehead{
\colhead{Parameter} & \colhead{Value}
}
\startdata
\hline 
\textbf{Solar Abundances} & \\
\hline
N(H) (cm$^{-2}$) & 1.32$\pm$0.05 $\times 10^{22}$ \\
T$_1$ (keV)     & 0.99$\pm$0.02 \\
v$_1$ (km s$^{-1}$)  & 738$\pm$90 \\
$\Delta$v$_1$ (km s$^{-1}$) & 1023$\pm45$ \\
EM$_1$ (cm$^3$) & 8.58$\pm0.44 \times 10^{56}$ d(kpc)$^2$ \\
Flux$_{1,abs}$ (erg cm$^{-2}$ s$^{-1}$) & 2.56$\pm 0.13 \times 10^{-11}$ \\
Flux$_{1,unabs}$ (erg cm$^{-2}$ s$^{-1}$) & 1.26 $\pm 0.06 \times 10^{-10}$ \\ 
T$_2$ (keV)  & 3.99$^{+0.22}_{-0.20}$ \\
v$_2$ (km s$^{-1}$)  & -522$^{+42}_{-54}$ \\
$\Delta$v$_2$ (km s$^{-1}$) & 1023$\pm45$ \\ 
EM$_2$ (cm$^3$) & 9.55$\pm0.50 \times 10^{56}$ d(kpc)$^2$ \\
Flux$_{2,abs}$ (erg cm$^{-2}$ s$^{-1}$) & 7.79$\pm 0.04 \times 10^{-11}$ \\
Flux$_{2,unabs}$ (erg cm$^{-2}$ s$^{-1}$) & 1.36$\pm 0.07  \times 10^{-10}$ \\
$\chi^2$  & 1.28 \\
\hline
\textbf{Variable Abundances} \\
\hline
N(H) (cm$^{-2}$)               & 8.3$^{+1.1}_{-1.5} \times 10^{21}$ \\
T$_1$ (keV)                    & 0.97$\pm0.03$ \\
v$_1$ (km s$^{-1}$)            & -867$^{+66}_{-81}$ \\
$\Delta$v$_1$ (km s$^{-1}$)    & 632$^{+146}_{-124}$ \\
EM$_1$ (cm$^3$)                & 0.06$^{+0.50}_{-0.06} \times 10^{56}$ d(kpc)$^2$ \\
Flux$_{1,abs}$ (erg cm$^{-2}$ s$^{-1}$)   & 1.98$^{+16.50}_{-1.98} \times 10^{-11}$ \\
Flux$_{1,unabs}$ (erg cm$^{-2}$ s$^{-1}$) & 6.20$^{+51.67}_{-6.20} \times 10^{-11}$ \\
T$_2$ (keV)                    & 4.06$^{+0.25}_{-0.23}$ \\
v$_2$ (km s$^{-1}$)            & -794$^{+150}_{-180}$ \\
$\Delta$v$_2$ (km s$^{-1}$)    & 1572$^{+180}_{-72}$ \\
EM$_2$ (cm$^3$)                & 9.7$^{+0.6}_{-0.5} \times 10^{56}$ d(kpc)$^2$ \\
Flux$_{2,abs}$ (erg cm$^{-2}$ s$^{-1}$)   & 8.77$\pm0.90 \times 10^{-11}$ \\
Flux$_{2,unabs}$ (erg cm$^{-2}$ s$^{-1}$) & 1.42$\pm0.16 \times 10^{-10}$ \\
$\chi^2$ & 1.12 \\
\hline
\text{Abundances} \\
\hline
[Ne/Ne$_\odot$] & 21$^{+143}_{-21}$      -- 10.0$_{-10.0}^{+3.0}$ \\
Mg/Mg$_\odot$   & 141$^{+160}_{-133}$    -- 1.0$\pm 1$ \\
Al/Al$_\odot$   & 172$^{+153}_{-63}$     -- 0.0$^{+1.0}$ \\
Si/Si$_\odot$   & 171$^{+89}_{-75}$      -- 1.5$\pm1.3$ \\ 
S/S$_\odot$     & 236$^{+319}_{-141}$    -- 1.8$\pm0.3$ \\
Ar/Ar$_\odot$   & 411$\pm232$            -- 1.6$\pm$0.6 \\  
Ca/Ca$_\odot$   & 969$^{+21}_{-500}$     -- 1.6$^{+0.6}_{-0.6}$ \\
Fe/Fe$_\odot$   & 110$^{+146}_{-96}$     -- 1.2$\pm$0.14 \\
\hline
\enddata
\end{deluxetable}
\begin{deluxetable}{rrrr}
\tablecaption{Flux of the emission lines: for the first group, H-like lines, we
 have obtained the flux by direct integration and subtracted the continuum measured
 in an adjacent spectral region without lines; for the triplets, for which we could
 not resolve the intercombination line, we have resorted to fitting three Gaussians
 with the same width and keeping the wavelength values shifted by the same amount
 for the r,i and f lines. The H-like lines and 
 the intercombination line are very closely spaced doublets 
 that cannot resolved. 
}
\tablehead{
\colhead{Ion} & \colhead{Rest $\lambda$}  &  \colhead{Observed  $\lambda$ }
 & \colhead{Flux $\times 10^{-13}$} \\
\colhead{}    & \colhead{(\AA)}      &  \colhead{(\AA)}
 & \colhead{(erg cm$^{-2}$ s$^{-1}$)}
} 
 \startdata
 S XVI  &  4.727 & 4.720 & 9.23$\pm$3.15 \\
   Si XIV  & 6.180 & 6.176  &  15.64$^{+1.91}_{-1.81}$ \\
   Al XIII & 7.171 & 7.165 & 1.18$\pm$0.59 \\
   Mg XII & 8.419  & 8.412 & 6.21$\pm$0.47 \\ 
   Ne X  & 12.123  & 12.119 & 5.4$_{-2.16}^{+0.97}$ \\
 S XV   r  & 5.039 & 5.032 & 7.73$^{+2.07}_{-1.82}$  \\
 S XV   i  & 5.063/5.067 &  & 2.12$^{+0.73}_{-1.09}$  \\
 S XV   f  & 5.10 & 5.101 & 5.14$^{+2.83}_{-1.99}$ \\
 Si XIII r & 6.648 & 6.639  & 7.71$^{+0.32}_{-1.96}$ \\
  Si XIII i & 6.685/6.688 &  & 4.13$^{+0.85}_{-0.79}$ \\
  Si XIII f & 6.740 & 6.730  & 5.19$^{+0.80}_{-0.96}$ \\
Mg XI r   & 9.169  & 9.155 & 2.36$^{+0.74}_{-0.63}$ \\
   Mg XI i   & 9.228/9.231  &  & 1.23$^{+0.27}_{-0.25}$ \\
   Mg XI f   & 9.314  & 9.300  & 0.85$^{+0.36}_{-0.30}$ \\
   Na X r & 11.003 &  10.954 &  2.06$^{+1.49}_{-1.08}$ \\
   Na X i    & 11.080/11.083 &  & 0.68$^{+1.12}_{-0.68}$ \\
   Na X f    & 11.192 & 11.143 & 0.83$^{+0.95}_{-0.62}$ \\
\enddata
 \end{deluxetable}
\begin{figure}
\epsscale{0.6}
\plotone{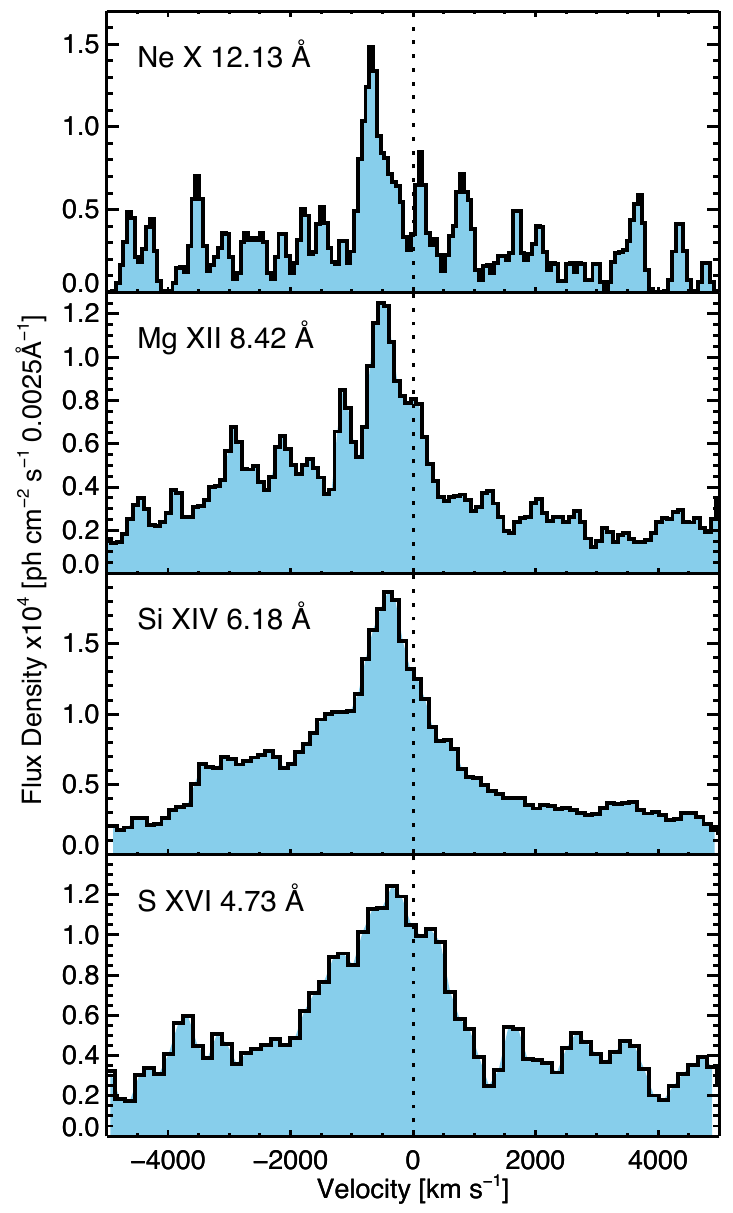}
\caption{The H-like lines
 of Ne X, Mg XII, Si XIV and S XVI  in the HEG spectrum,
 plotted in velocity space, showing the asymmetric, blue-shifted profiles.}
\label{fig:Jeremy}
\end{figure}
\begin{figure}
\epsscale{0.6}
\plotone{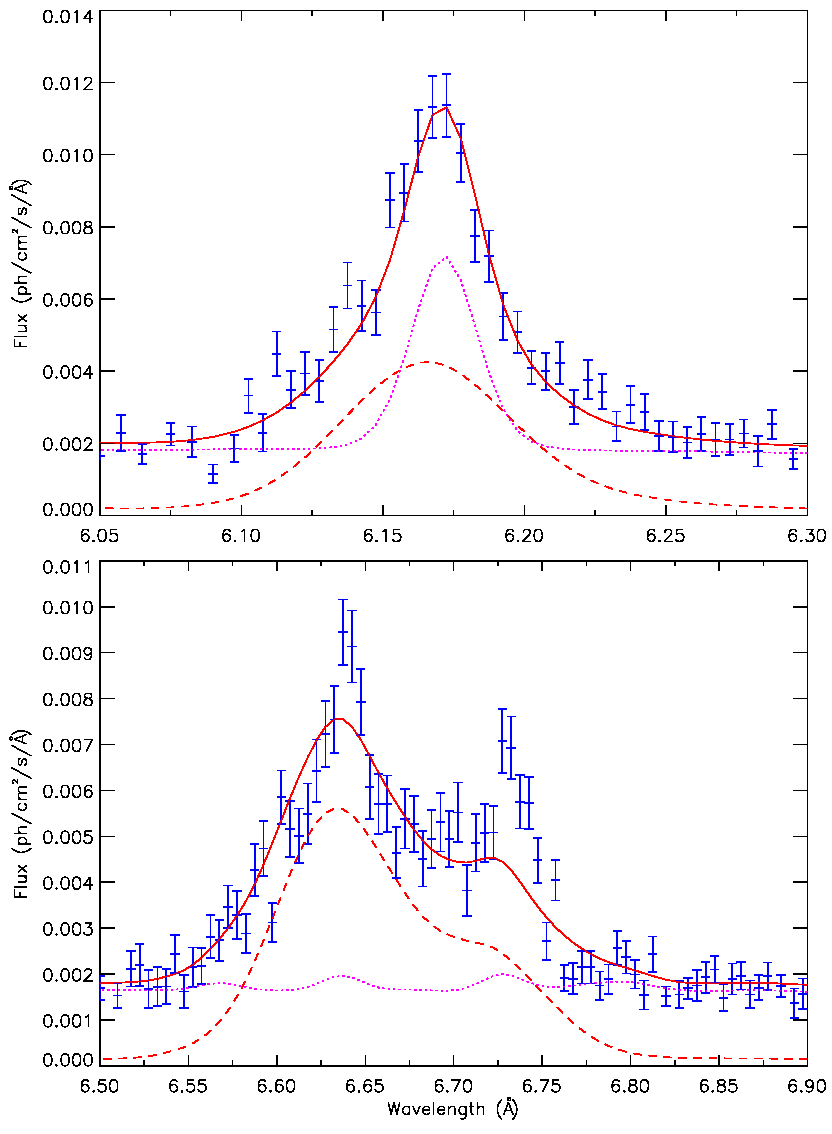}
\caption{Upper panel: the Si XIV H-like line 
 in the MEG spectrum, binned by signal-to-noise $\geq$20
 as fitted with two components with the second
 model in Table 2. The pink dotted line shows 
 the hotter component ($\simeq$ 3.5  keV), the red dashed line the
 $\simeq$ 1 keV component, and the coadded final model is plotted with a thick
 solid red line. Lower panel: 
 the same for the Si XIII He-like triplet as observed with the MEG,
 also with a binning by S/N=20: here 
 the hotter component, traced
 by the pink dotted line, does not contribute to the lines, it only
 enhances the continuum level.}
\label{fig:global-Gaussians}
\end{figure}
\begin{figure}
\epsscale{0.6}
\plotone{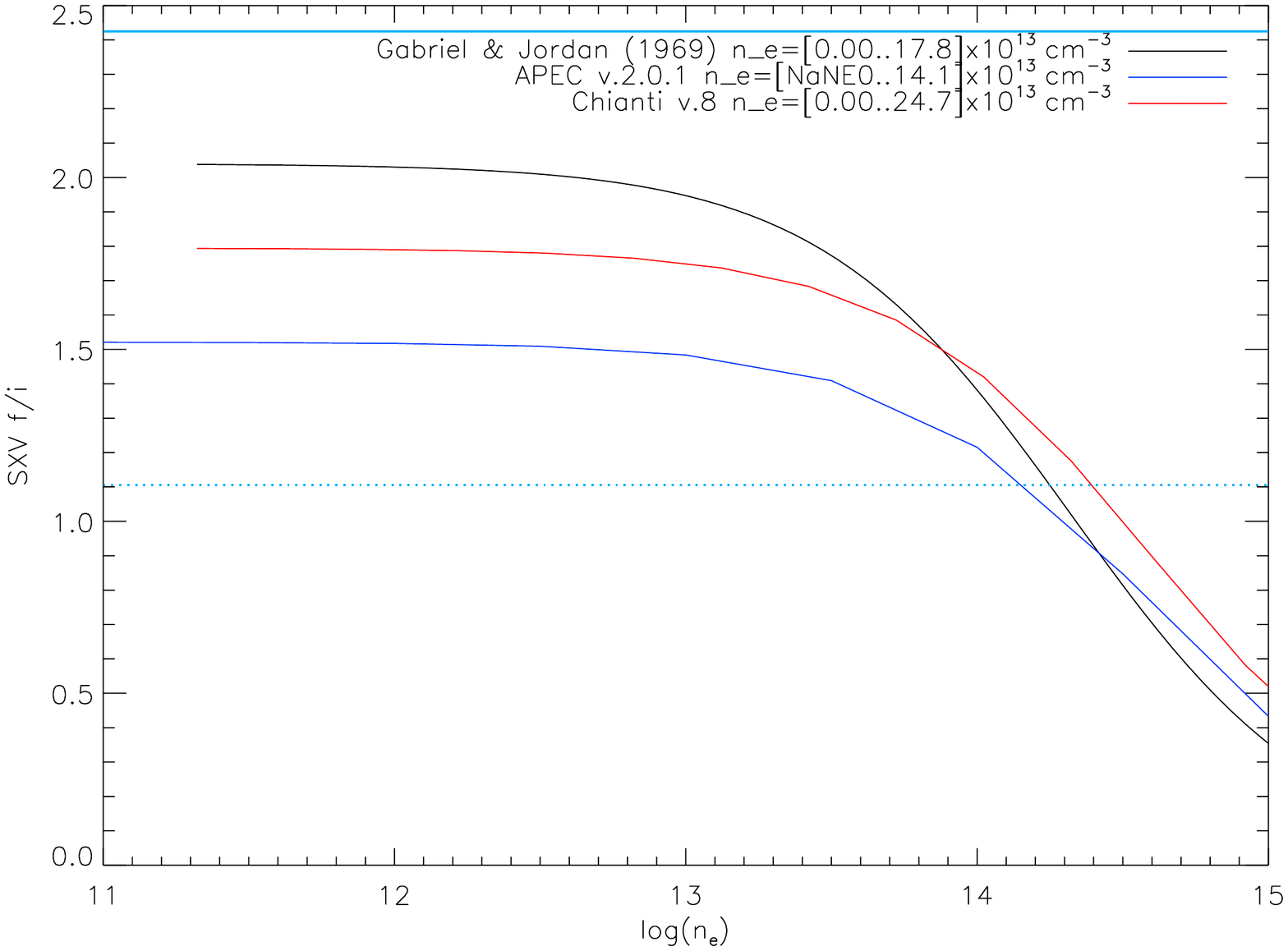}
\plotone{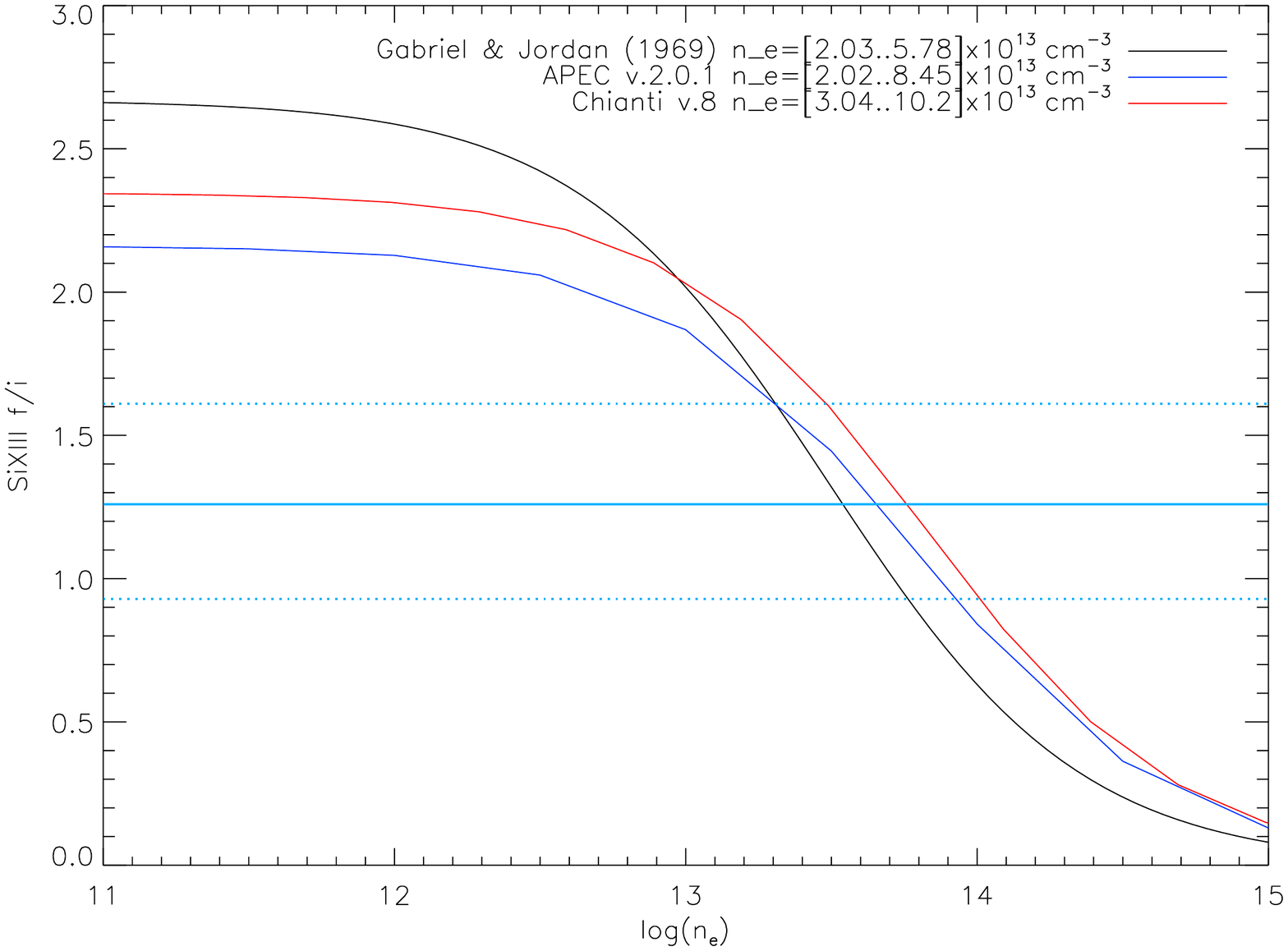}
\plotone{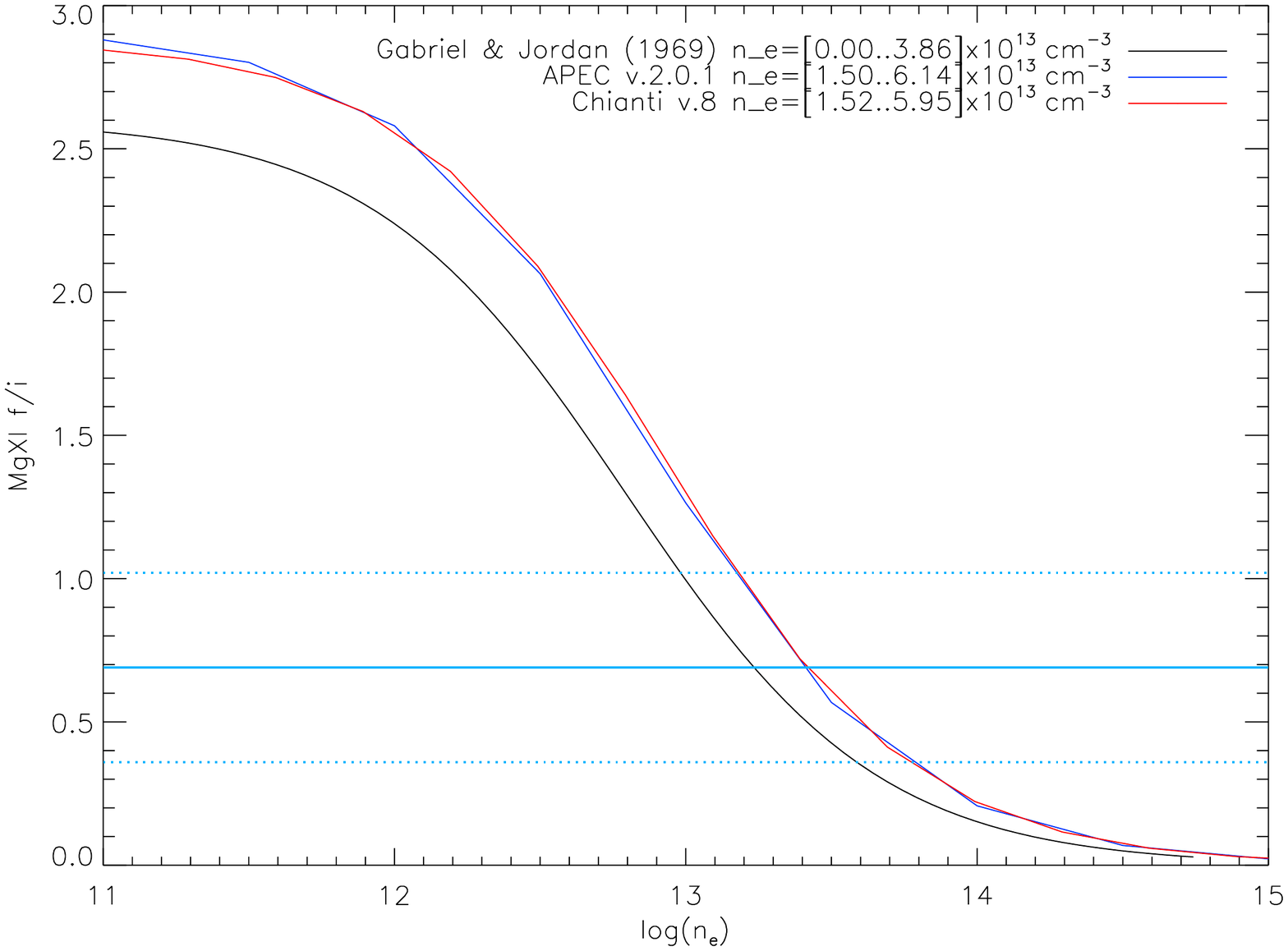}{}
\caption{The f/i ratio of the S XV, Si XIII and Mg XI triplets
 as a function of density in absence of strong photoexcitation, according
 to the \citet{Gabriel1969}, APEC \citep{Smith2001}
 and Chianti \citep{DelZanna2015} databases. The solid blue 
 line shows the measured value, the dotted lines show the lower
 and upper limit
 given the 1$\sigma$ uncertainty (only the lower limit for S XV).}
\label{fig:januwe}
\end{figure}

\section{Comparison with the Chandra HETG spectra of other novae}
The known physical parameters of V3890 Sgr have been compared with the
 other Galactic symbiotic RNe in Table 1.
A comparison of our spectra with
 those obtained in the early phases of other symbiotic
 RNe outbursts with the same gratings 
 shown in Fig. 7, indicates rather remarkable similarities
 between this nova, the two other recurrent symbiotic novae
 that were observed with X-ray gratings, and the classical nova V959 Mon, which was also
 observed before the supersoft X-ray phase, when the central source
 outshines the spectrum of the ejecta.
 The HETG
 spectra of RS Oph were presented by \citet{Nelson2008, Ness2009},
 the one of V745 Sco by \citet{Drake2016}, and that of V959 Mon 
 by \citet{Peretz2016}. Table 4 lists the X-ray flux,
 absorbed and unabsorbed X-ray luminosity of these novae when their ejecta were observed
with X-ray gratings. 
 V745 Sco had an extremely short SSS phase and the central
 source was contributing to the X-ray flux copiously on day 10 in
 the {\sl Swift} XRT range, so the actual flux from the ejecta is
 between the value derived from {\sl Swift+NuSTAR} and {\sl NuSTAR} alone 
 (3-79 keV range). When it was observed with the {\sl Chandra}
HETG on day 14, the central source had already turned off.  

A a distance of 4.4 kpc, which is likely to be only a lower limit,
 V3890 Sgr on day 7 had a large X-ray luminosity,
 L$_{\rm X}=1.91 \times 10^{35}$ erg s$^{-1}$, and the unabsorbed
 luminosity was a factor of 2 to 3 higher. This X-ray luminosity
 is also comparable to that of RS Oph and V745 Sco. 
Like the gamma-ray luminosity, the X-ray luminosity due to
 the shocked ejecta also seems to be much larger for the symbiotic novae than
 for the short period ones. Short period novae in outburst before
 (or after) the supersoft X-ray phase do not exceed 
 an X-ray luminosity of 10$^{33}$ erg s$^{-1}$ \citep{Orio2001,
 Schwarz2007}. Since symbiotic novae
 differ from the others because of the red giant companion,
this is considered an indication
 that the impact with the red giant wind must be causing the shocks.  
 An exception is the classical nova
 V959 Mon, which has a 7 hour orbital period and is not a symbiotic. 
 It exploded when it was close to the Sun and could not
 be observed until much later, so the outburst day 
 of the HETG observation is an estimate. This nova
 was discovered as a gamma-ray source with Fermi, and 
 it may have been much more X-ray luminous previously. However, 
 the hard X-ray luminosity about two months after the 
 estimated optical maximum was 
 still very high. If V959 Mon is not a RN (no
 previous outburst was ever reported), it is an exception among short
 period novae.  This nova would have had time to accrete a larger
 envelope than  RNe,  ejecting a much larger mass that 
 slowed the X-ray flux emergence and evolution,
 but it is difficult to explain
 the high X-ray luminosity without a new episode of mass ejection
 after the initial peak.

 The spectra of V745 Sco and RS Oph differ from that of V3890 Sgr in
 the soft portion, where no significant emission lines of oxygen
 and nitrogen were measured. The APEC fits to the RS Oph and V745 Sco
 spectra in fact required more than only two components to explain
 the observed range of atomic transitions \citep{Nelson2008, Drake2016}. 
 An important difference between V959 Mon and the symbiotic RNe
 are the ratios of lines of different species. 
The ratios of emission lines that
 are the strongest in oxygen-neon (ONe) novae, such as V959, indicate
 nucleosynthesis and elemental
 abundances predicted to be the result of a TNR on an ONe WD,
 where  Ne-Na and Mg-Al cycles, that do not occur
on CO WDs, operate in addition to the CNO cycle 
\citep{Jose1998, Starrfield2009, Kelly2013}. Atomic
 Mg (sum of three isotopes) is also overabundant
 in a superficial layer on ONe WDs.
 The emission lines of Si and Fe are stronger than those of Mg, Ne and Al
  in our spectrum of V3890 Sgr. V959 Mon instead has more abundant
 flux in the Mg, Ne and Al lines than in those of Si and
 Fe, even if the spectrum was fitted with two plasma components at
 approximately the same temperature we obtained in this paper 
 for V3890 Sgr \citet{Peretz2016}. If we compare V3890 Sgr to RS Oph
 and V745 Sco, the line flux ratios of
 the three symbiotic RNe spectra are instead  about the same. 
 Of course, if the material has mixed
 with the red giant wind, we do not
 expect to find ``clean'' signatures of the ashes burned on the WD, so
 a conclusion on the WD nature (whether it is of CO or ONe) does
not always seem feasible for symbiotic-RNe on the basis of the
 X-ray emitted in the nova outflow, but the difference 
 with the V959 Mon spectrum is rather striking.  
\subsection{Cooling time and X-ray flux at quiescence}
 A discussion of the cooling time of the ejecta in post-outburst
 symbiotic RNe is found in \citet{Moore2012}. These authors 
 assume a spherical explosion and that shocks occur typically 
 at a distance of few AU from the
 red giant and discuss the implications for a symbiotic RN ending
 its life in a type Ia supernova explosion. 
 X-ray flux after the RN outburst 
 should only be detectable for a ``cooling time'' found to
 vary from few days to few weeks. It is predicted to be about 
 2 weeks for RS Oph, and after this time the kinematics of the ejecta are dominated by
 momentum (rather than energy) conservation. 

 \citet{Sokoloski2006} observed that the  X-ray flux from the ejecta 
 of RS Oph was related to the time t in days since maximum as t$^{-5/3}$. 
 The {\sl Swift} XRT archival observations 
 of V3890 Sgr show that V3890 Sgr  cooled 
 much more slowly than RS Oph, since the flux
 on 2019 November 19 was still 6.7 $\times 10^{-12}$ erg cm$^{-2}$ s$^{-1}$
 in the 0.3-10 keV range of the {\sl Swift} XRT.

Another interesting measure is the X-ray flux at quiescence, and
 whether it originates in the boundary layer of
 an accretion disk and gives a  measure of the accretion rate.
 After the 1990 outburst, V3890 Sgr was still
 detected in X-rays in the short {\sl ROSAT} All-Sky-Survey exposures 
\citep{Orio1993} about 5 months after maximum, but it was
 no longer detected in serendipitous pointings 11 and 17 
 months after the outburst \citep{Orio2001}. This is
 consistent also with an {\sl XMM-Newton} exposure on
2010 April  (archival observation by P.I. Sokoloski) in which we
 found an upper limit for the X-ray luminosity at 4.5 kpc of
 approximately 10$^{31}$ erg s$^{-1}$. 
V745 Sco, for comparison,  was only marginally 
 detected four years before the last outburst with {\sl XMM-Newton},
 with   X-ray luminosity 6 $\times 10^{31}$ erg s$^{-1}$ 
 in the 0.3-8.0 keV range \citep[see][]{Luna2014}.  
\begin{figure}
\plotone{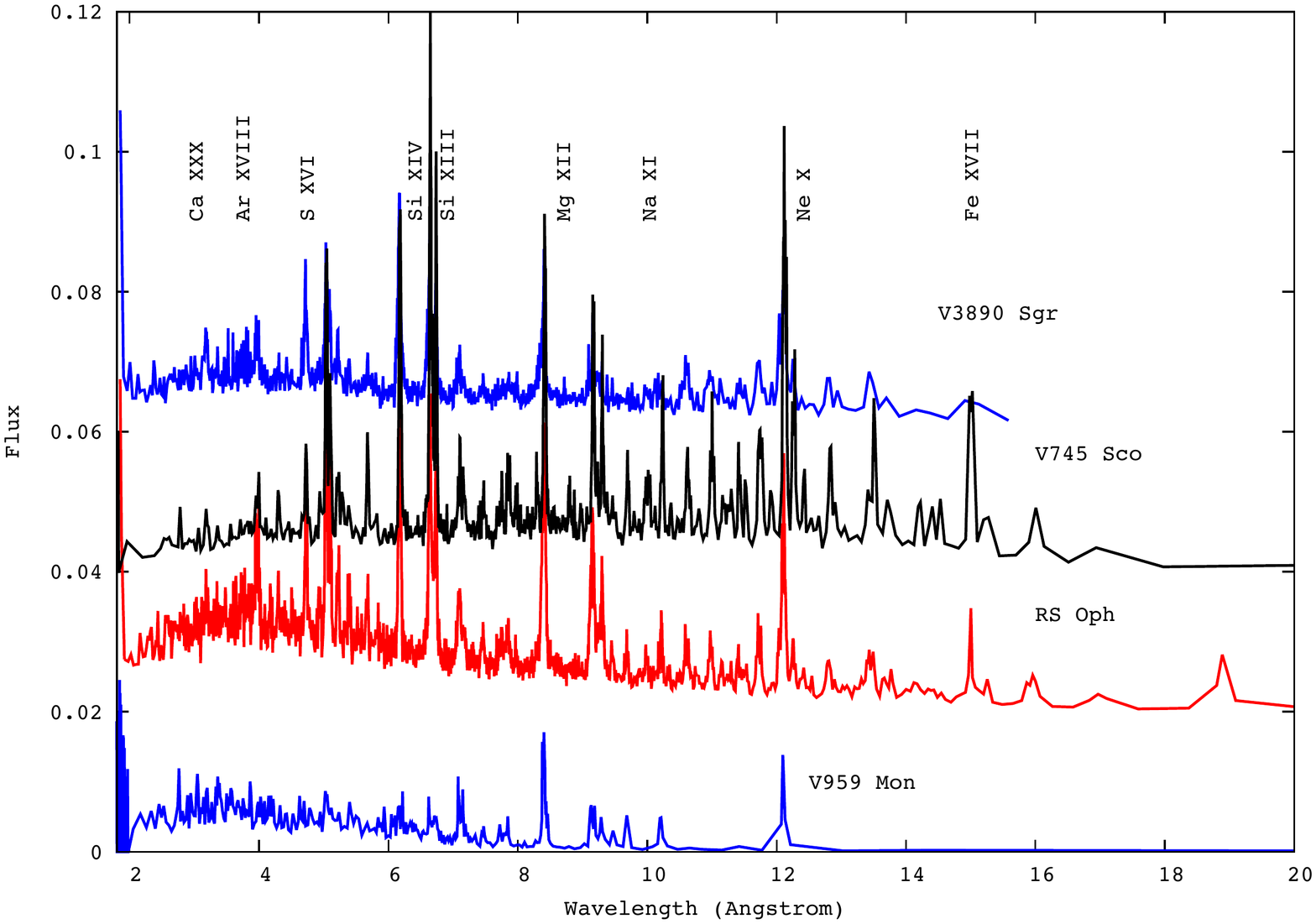}
\caption{Comparison of the MEG spectra of three symbiotic recurrent novae,
 V3890 Sgr included, and of the classical nova V959 Mon.
 The flux on the y  axis
 is measured in units of photons \AA$^{-1}$ cm$^{-2}$ s$^{-1}$,
 but for  V745 Sco and V959 Mon it has been multiplied by a
 factor of 10; moreover the spectra above V959 Mon have
 each been offset, from bottom
 to top, by 0.02 photons  \AA$^{-1}$ cm$^{-2}$ s$^{-1}$
with respect to the one below it, for clarity. 
}
\end{figure}
\begin{deluxetable*}{rrrrr}
\tablecaption{X-ray emission of the ejecta of three symbiotic RNe and V959 Mon.
 The assumed distances are those in boldface in Table 1.
The flux was integrated by us using the archival observations,
 for V745 Sco on day 10 it is reported from Orio et al. (2015; Swift 
 was also used, but it  included the strong SSS emission, so
 only the NuSTAR flux is given here in the 3-79 keV range,
 to isolate the central emission) and
 for RS Oph on day 7 it is from \citet{Sokoloski2006} in
  the 3.0-20 keV range of RossiXTE.  We regard these as upper limits
 to the total flux of the ejecta.  For V959 Mon the flux
 is in the 0.3-10 keV band.}
\tablehead{
\colhead{Nova} & \colhead{Day} & \colhead{Flux} & Luminosity & Lum. (unabs.) \\
               &               & \colhead{(erg cm$^{-2}$ s$^{-1}$)} &
\colhead{(erg s$^{-1}$)} & \colhead{(erg s$^{-1}$)}   
}
\startdata
V3890 Sgr & 8 & 1.01 $\times 10^{-10}$ & 2.33 $\times 10^{35}$ & 4.6-7 $\times 10^{35}$ \\
RS Oph    & 7 &  $\geq$1.98 $\times 10^{-9}$   & $\geq$7 $\times 10^{35}$ & \\
          & 14 & 6.09 $\times 10^{-10}$ & 1.86 $\times 10^{35}$ & \\
V745 Sco  & 10 & (1.6$<$F$<$2.39) $\times 10^{-10}$  & (1.16$<$L$<$1.73) $\times 10^{36}$ &  \\ 
          & 16 & 4.21 $\times 10^{-11}$ & 3.05 $\times 10^{35}$ & \\ 
V959 Mon  & 83 & 1.7  $\times 10^{-11}$ & 2.33 $\times 10^{34}$ & \\ 
\enddata
\end{deluxetable*}

  The X-ray flux of RS Oph at quiescence is variable,
  and on average it seems to have increased between 1991-1992 and 2007-2008
\citep{Orio1993, Orio2001, Nelson2011}. \citet{Orio1993} found
 that the X-ray flux measured in 1991 was not consistent with
 the boundary layer of a disk with high $\dot m$,
 but an observation in April of 2008 with
 Chandra showed an X-ray flux that was marginally consistent
 with $\dot m \geq 2 \times 10^{-8}$ M$_\odot$ yr$^{-1}$ 
\citep{Nelson2011}, explaining the occurrence
 of outbursts every 10-20 years with model calculations 
 with the same $\dot m$ value \citep[e.g.][]{Yaron2005}.

Only T CrB, which is at less than 1 kpc distance, emits significant,
 variable X-ray flux at quiescence that can  be
 attributed to accretion \citep[see][]{Luna2008, Luna2013,
Luna2018}. It is also quite a
 variable and hard X-ray source, as expected for a disk
 boundary layer around a massive WD \citep{Ilkiewicz2016, Luna2018}.
 T CrB stands out, in comparison
 with the other symbiotic RNe, with its conspicuous hard
 X-ray flux at quiescence. It has never been observed in X-rays in outburst,
 and the next eruption would be an opportunity not to be missed. 
\section{Discussion and conclusions}
On the 7th  day after the optical maximum, V3890 Sgr was emitting an 
X-ray luminosity of 4.6-7 $\times 10^{35}$ erg s$^{-1}$ in the
 0.7-9 keV range (depending on
 absorbing column density, and assuming a distance
 of 4.5 kpc, which is likely a lower limit). We attribute this emission to
 the  plasma outflowing from the nova, and suggest that strong 
shocks formed when it
 impacted the circumbinary medium  filled by the red giant
 wind.  The same physical explanation has been suggested for other symbiotic
 novae, that typically emit much higher X-ray flux than short period
 systems.  The spectrum of V3890 Sgr
 can be fitted with a model of thermal plasma in CIE, and shows
 prominent,
 asymmetric emission lines of Mg, Ne, Na, Si, S, Ar, Ca and Fe.

Our analysis provides the following important points:

1) The lines due to H-like transitions are stronger than the lines
 due to He-like transitions. This can be explained by at least 
 two thermal plasma components, and in
 fact we obtain a fit to the spectrum with
 two components in collisional ionization equilibrium,
 at temperatures of  approximately
 1 keV and 4 keV, respectively.
 The hotter plasma is  almost completely ionized
 and does not contribute to the emission lines due to He-like transitions. 
 A possible interpretation is that the two components represent 
 a forward and a reverse shock.

2) An alternative interpretation can be suggested by analogy
with a model for RS Oph of \citet{Orlando2009}: 
 the two plasma components may represent, respectively,
 colder material that is not much mixed with the red giant wind,  and
 hotter material that is instead heavily mixed and has been heated. 
In our model fit,
 there is in fact some evidence that the cooler plasma has enhanced
 metal abundances with respect to the hotter component, resembling 
 two regions   
 in the hydrodynamical model of \citet{Orlando2009} for RS Oph,
which predicts that the zones containing
 mostly ejecta material are  denser and colder, while 
 in regions where shocked ejecta have mixed with circumstellar material
  the plasma is hotter and dominated by thermal conduction.
For RS Oph, the authors were able to model the He-like resonance line profiles
 with the overlap of the two components,  in a similar manner
 as in our Fig. 5 for the H-like lines. 
  Applying  this interpretation to
 V3890 Sgr, the higher abundances of the cold region
 indicated by the model with free abundances may be due
 to nova nucleosynthesis products in the ejecta.
 The  hotter, mixed ejecta on day 5 for RS Oph would have contributed
 to 80\% of the observed flux, which is similar to with our finding that the
 hotter plasma for V3890 Sgr on day 6  contributed to at least 70\%
 of the observed flux.

3) Our {\sl APEC} spectral model in {\sl XSPEC}
fits the data, but it is approximate and phenomenological, while 
 only a detailed hydrodynamical model may quantify the effect
  of the differential absorption
 and disentangle the exact contribution of the two (or possibly more) different
 plasma regions to the line profiles.  One reason for which the 
 model is still too simplistic is that we have only one absorbing component with
 characteristics corresponding to the ISM along the line of sight.  
 Instead, there also is significant intrinsic absorption of the nova
 ejecta (in fact, the column density was observed with the {\sl
 Swift XRT to decrease as the ejecta expanded; Page et al. 2020, 
 in preparation). We suggest that only a detailed
 hydrodynamical model may explain the formation of the different
 emission lines in a rigorous
 way and account for profile and flux of each single line.} 
 
4) The X-ray spectrum of V3890 Sgr differs from that of the two
 symbiotic RNe with similar orbital parameters, RS Oph and V745 Sco,
 in the line ratios of the He-like triplets.  The $R=f/i$ line
 ratios are quite smaller than in the spectra of the two
 previous symbiotic RN.  The values we evaluated indicate 
 high electron density (a few 10$^{13}$ cm$^{-3}$), 
 unless there was still significant UV flux from the photosphere.
 Unfortunately, at the epoch the spectrum was taken, 
it is difficult to establish
 whether the emission of the WD photosphere was already 
 peaking in the extreme UV or in the supersoft X-rays.  
 If the density was indeed very high, the emission must have arisen
 from very clumpy ejecta constituting only a small fraction of
 the emitted material, of order  10$^{-12}$ M$_\odot$.
  This is not a completely unusual and new scenario; in fact
 similarly high electron density has been
 inferred for other novae \citep{Orio2013, Tofflemire2013, Peretz2016},
 suggesting that in some novae the shocked material that emits X-rays  
 may be
emitted as dense ``bullets'', in a clumpy rather than smooth outflow.

5) There are interesting caveats relative to the possibility
 of very high electron density. First of all,  the cooling time of the
 shock would be much, much shorter than in the models of \citet{Orlando2009, Moore2012}
 for RS Oph. It would be  only about 1 s, so the shock must be
 continuously powered, requiring a constant mass outflow for
 a prolonged time.
 Second, and most important, in case of non-radiative shocks, 
 the density ratio of pre- and post-shock material
 should be about 4.  Even taking into account radiative cooling,
 the density gradient
 cannot have been as large as orders of magnitude, however
 the typical red giant wind electron density at $\simeq$4 AU from
 the red giant (where \citet{Orlando2009}
 estimate that the shocks occurred in the RS Oph case) 
 is only 10$^9$ cm$^{-3}$ \citep[see analytical formula in][]{Moore2012}. 
 Even with the proposed equatorial density enhancement, the density is
 only of the order of 10$^{10}$ cm$^{-3}$. 
 Since the density scales with the square of the distance from the red giant,
 perhaps the shocks in V3890 Sgr did not occur at a distance
 of a few AU from the red giant,  but the 
 site of X-ray emission was instead very close to the giant's photosphere. 
 In this case, the emitting region subtended a small solid angle
 and occupied a small volume. 

 With a spectrum taken at only one epoch, we cannot distinguish between the
 two possibilities: emission originating at distance from
 the star of a few AU, which implies that the $R$ line ratio is  the
 product of a strong photoionizing source, and an actual 
 electron density of order of 10$^{13}$ cm$^{-3}$. 
  We know that two days after our spectrum was taken, the photoionizing flux
 was negligible, but
 the only other high resolution X-ray spectrum of V3890 Sgr 
 was obtained only 11 days later \citep[][and Ness et al. 2020,
 in preparation]{Ness2019} and
 unfortunately this second spectrum was taken too late.
 Although at this epoch the central supersoft X-ray source emitted negligible
 UV flux and did not photoexcite of the $i$ line,
  the flux had decreased and measurements of the $R$ ratios
 were not feasible.
 This highlights the fact that, in a fast symbiotic nova like
 V3890 Sgr, the cadence of the spectral exposures should be
 evaluated carefully in the future: multiple exposures, close in time, would be
 very useful. 
 A second spectrum taken only few days later, when the central source
 was peaking in the X-rays and already clearly
 not contributing significant UV flux, may have solved the conundrum.

6) Even if  we have highlighted this intriguing possibility that
the V3890 Sgr spectrum may have originated in shocks
 within a small volume of dense emitting plasma, close to the
 red giant photosphere, instead of a site at a distance of a few AU from
 the red giant like in RS Oph and V745 Sco, the 
X-ray luminosity of V3890 Sgr 
 is in the same range observed for the two other
 symbiotic RN observed in hard X-rays in the two weeks following the optical
 maximum, and it is about two orders of magnitude higher than in most 
 short period classical novae.
  Another symbiotic nova, V407 Cyg, not known to be recurrent, also had very 
 X-ray luminous ejecta \citep{Orlando2012}.
 Only the short period nova V959 Mon was observed
 to have an X-ray luminosity of a few 10$^{34}$ erg  s$^{-1}$    
 3 months after optical maximum; all other short period
 novae have much less X-ray luminous ejecta. The early X-ray spectra of all three
 symbiotic RN observed so far with the X-ray gratings show remarkable
 similarities not only in the luminosity level,
 but also in line profiles, and (from initial estimates based
 on the line fluxes) probably in the chemical abundances of the emitting material.

7) The observed thermal X-ray emission cannot be directly
 linked to the measured gamma-ray flux for this nova.
  The gamma-rays would originate in the initially X-ray shocked plasma only
 if the thermal plasma was much hotter, and if there was
 also the contribution of a non-thermal component to the X-ray flux 
\citep{Metzger2015}.  However, we would like to note that
 the possibility  of a dense and clumpy plasma evoked
 by the high electron density gives some support to the idea that also
  possible initial shocks giving rise to the gamma-rays
  may be
 difficult to observe, because they would easily be ``buried'' within
 a large volume of non X-ray emitting, absorbing material \citep[as suggested
 by][]{Nelson2019}.

8) The relative ratios of the flux in Al, Mg and Ne lines (typical of
 novae on ONe WDs) to the flux in lines of Si and Fe
 (which are not related specifically to ONe WDs),
  in the spectrum of V3890 Sgr, like in the other symbiotic
 RNe RS Oph and V745 Sco,
 are much lower than in a known ONe short period nova, V959 Mon.
 This suggests that V3890 Sgr hosts a CO WD.

 In the future, obtaining high resolution
 emission line spectra of the ejecta of symbiotic RNe within a few
 days, should allow us to assess the possibility of high electron density,
 with all its implications {for the emission
 site, emission measure, and the inferred clumpiness in
 the outflow.  Perhaps in V3890 Sgr a first mechanism of 
 shocks production in a small volume close to
 the red giant evolved into shocks occurring farther from
 the red giant at a later epoch, and in a medium of much lower density,
 as inferred in the observations of V745 Sco and RS Oph,
  and this may also explain the longer than predicted cooling time
 estimated with the {\sl Swift} XRT. 
 Further research is also encouraged in modeling our existing
 data, e.g. with a detailed hydrodynamical calculations to seek the
 most appropriate physical model. 
 
 High resolution X-ray
 spectroscopy of nova ejecta has a rich potential to lead to discoveries
 in nova physics. We suggest these
 observations should be done whenever this is possible,
 to widen the database and extend it also to non-symbiotic 
 and classical novae. We foresee that the Jaxa/NASA/ESA X-Ray Imaging
 and Spectroscopic Mission {\sl XRISM} \citep[see][]{Williams2019} will produce 
 results with even higher S/N, and 
 later {\sl Athena} \citep[see][]{Dandrea2019} will allow much shorter exposure times,
 thereby extending this type of work to many more novae at repeated epochs, to
 make comparison with outburst properties at other
 wavelengths and construct a coherent picture of the challenging puzzle
 of the nova physics. 
\acknowledgments
We thank Dina Prialnik for sharing unpublished results with us.
M.O. and J.D. acknowledge support of a Chandra award to
work on this observation, obtained as Director Discretionary
 Target of Opportunity. We are grateful to Belinda Wilkes,
 Chandra X-Ray center director, for scheduling this observation
 on short notice.  Other team
 members acknowledge support from the following funding agencies:
 E.B. by a Center of Excellence of THE ISRAEL SCIENCE FOUNDATION (grant No. 2752/19);
G.J.M.L, as a member of the CIC-CONICET (Argentina),
 from grant PICT 0901/207; J.M.  from the National Science Centre, Poland,
through grant OPUS 2017/27/B/ST9/01940; N.P.M.K. and
 K.L.P. from the UK Space Agency; S.S. from NASA support to ASU;
M.J.D. from the UK Science and Technology Facilities Council (STFC); R.D.G. from
NASA and the United States Air Force (USAF); C.E.W. from NASA.

\bibliography{mybib}{}
\bibliographystyle{aasjournal}
\end{document}